\documentclass[aps,prd,floatfix,nofootinbib,showpacs,twocolumn,superscriptaddress]{revtex4-1}

\usepackage{mathrsfs}
\usepackage{amsmath}
\usepackage{amssymb}
\usepackage{bbold}
\usepackage{color}
\usepackage{graphicx}
\usepackage{soul}

\definecolor{purple}{rgb}{0.5,0,0.5}
\definecolor{blue}{rgb}{0.0,0,0.9}

\begin{document}
\title{Critical endpoint  in the presence of a chiral chemical potential}

\author{Z.-F. Cui}
\email{phycui@nju.edu.cn}
\affiliation{Department of Physics, Nanjing University, Nanjing, Jiangsu 210093, China}

\author{I.\,C. Clo\"{e}t}
\email{icloet@anl.gov}
\affiliation{Physics Division, Argonne National Laboratory, Argonne,
Illinois 60439, USA}

\author{Y. Lu}
\email{luya@smail.nju.edu.cn}
\affiliation{Department of Physics, Nanjing University, Nanjing, Jiangsu 210093, China}

\author{C.\,D. Roberts}
\email{cdroberts@anl.gov}
\affiliation{Physics Division, Argonne National Laboratory, Argonne,
Illinois 60439, USA}

\author{S.\,M. Schmidt}
\email{s.schmidt@fz-juelich.de}
\affiliation{Institute for Advanced Simulation, Forschungszentrum J\"ulich and JARA, D-52425 J\"ulich, Germany}

\author{S.-S. Xu}
\email{xuss@nju.edu.cn}
\affiliation{Department of Physics, Nanjing University, Nanjing, Jiangsu 210093, China}

\author{H.-S. Zong}
\email{zonghs@nju.edu.cn}
\affiliation{Department of Physics, Nanjing University, Nanjing, Jiangsu 210093, China}
\affiliation{Joint Center for Particle, Nuclear Physics and Cosmology, Nanjing, Jiangsu 210093, China}

\date{28 April 2016}

\begin{abstract}
A class of Polyakov-loop-modified Nambu--Jona-Lasinio (PNJL) models have been used to support a conjecture that numerical simulations of lattice-regularized quantum chromodynamics (QCD) defined with a chiral chemical potential can provide information about the existence and location of a critical endpoint in the QCD phase diagram drawn in the plane spanned by baryon chemical potential and temperature.
That conjecture is challenged by conflicts between the model results and analyses of the same problem using simulations of lattice-regularized QCD (lQCD) and well-constrained Dyson-Schwinger equation (DSE) studies.
We find the conflict is resolved in favor of the lQCD and DSE predictions when both a physically-motivated regularization is employed to suppress the contribution of high-momentum quark modes in the definition of the effective potential connected with the PNJL models and the four-fermion coupling in those models does not react strongly to changes in the mean-field that is assumed to mock-up Polyakov loop dynamics.
With the lQCD and DSE predictions thus confirmed, it seems unlikely that simulations of lQCD with $\mu_5>0$ can shed any light on a critical endpoint in the regular QCD phase diagram.
\end{abstract}


\maketitle


\noindent\textbf{I.$\;$Introduction}.
\label{intro}
One of the most basic questions in the Standard Model refers to unfolding the state of strongly-interacting matter at extreme temperature and density: the former existed shortly after the Big-Bang and the latter is thought to exist in the core of compact astrophysical objects.  Quantum chromodynamics (QCD) is supposed to provide the answer, which hinges on the existence and interplay between color confinement and dynamical chiral symmetry breaking (DCSB), two emergent phenomena whose domains of persistence and disappearance characterize a potentially very rich phase structure.  Confinement is most simply defined empirically: those degrees-of-freedom used in defining the QCD Lagrangian (gluons and quarks) do not exist as asymptotic states, \emph{i.e}.\ these partonic excitations do not propagate with integrity over length-scales that exceed some modest fraction of the proton's radius.  The forces responsible for confinement appear to generate more than 98\% of the mass of visible matter \cite{national2012NuclearS, Brodsky:2015aia}.  This is DCSB, a quantum field theoretical effect that is expressed and explained via, \emph{inter alia}, the appearance of momentum-dependent mass-functions for quarks \cite{Bhagwat:2003vw, Bowman:2005vx, Bhagwat:2006tu, Roberts:2007ji} and gluons \cite{Cornwall:1981zr, Aguilar:2008xm, Boucaud:2011ug, Ayala:2012pb, Binosi:2014aea, Aguilar:2015bud}, and helicity-flipping terms in quark--gauge-boson vertices \cite{Chang:2010hb, Bashir:2011dp, Qin:2013mta, Rojas:2013tza, Aguilar:2014lha, Mitter:2014wpa}, all in the absence of any Higgs-like mechanism.

Owing to the complexity of strong interaction theory, attempts are often made to develop insights concerning confinement, DCSB, and the associated phase diagram in the plane spanned by quark chemical potential ($\mu$) and temperature ($T$) by using simple, tractable models.  The properties and predictions of one such class of models are the subject of our analysis; namely, the Polyakov-loop-modified Nambu--Jona-Lasinio (PNJL) models \cite{Fukushima:2003fw}, which introduce a mock-up of color confinement into the Nambu--Jona-Lasinio (NJL) model through the expedient of a static potential whose behavior is tuned to emulate Polyakov loop dynamics \cite{Svetitsky:1985ye}.

Chiral symmetry restoration in QCD is a second order transition in the chiral limit at nonzero temperature and small chemical potential.  This transforms into a crossover at realistic values of the current-quark masses; and numerous analyses suggest that it becomes a first-order transition when the chemical potential exceeds a certain minimum value, so that a critical endpoint (CEP$_\chi$) should be a salient feature of the phase diagram \cite{Stephanov:2007fk}.  Although the existence and location of CEP$_\chi$ is currently both a model-dependent statement, as reviewed, \emph{e.g}.\ in Refs.\,\cite{Qin:2010nq, Fischer:2014ata, Ayala:2014jla}, and a problem that is intractable using contemporary lattice-QCD (lQCD) algorithms \cite{Aarts:2015tyj}, an experimental search is underway \cite{Kumar:2012fb, Soltz:2014dja}.


In connection with theoretical analyses aimed at locating CEP$_\chi$, it has been conjectured that numerical simulations of lQCD defined with a chiral chemical potential, $\mu_5$, which can be performed without complications \cite{Yamamoto:2011gk}, may serve as a surrogate for simulations with $\mu\neq 0$, insofar as a critical endpoint in the $(\mu=0,\mu_5,T)$-plane, CEP$_5$, entails the simultaneous existence of CEP$_\chi$ in the $(\mu,\mu_5=0,T)$-plane and might also provide a means of determining the approximate location of CEP$_\chi$ \cite{Ruggieri:2011xc}.  The argument was supported therein by results obtained using a PNJL model.  Notably, a CEP$_5$ is also located in other models with similar qualitative features \cite{Fukushima:2010fe, Chernodub:2011fr, Gatto:2011wc} and $\mu_5>0$ was typically found to decrease the temperature associated with chiral symmetry restoration: $T^\chi_{\mu_5>0} < T^\chi_{\mu_5=0}$.

Taking this suggestion seriously, lattice simulations were performed at $\mu_5\neq 0$, with a surprising outcome, \emph{viz}.\ no CEP$_5$ was found and, moreover, $T^\chi_{\mu_5>0} > T^\chi_{\mu_5=0}$ \cite{Yamamoto:2011gk, Braguta:2015zta, Braguta:2015owi}.  Both results contradict the model studies.  In another curious twist, the lQCD results were confirmed in studies \cite{Wang:2015tia, Xu:2015vna} that produced solutions of the dressed-quark gap equation at $(\mu,\mu_5,T)>0$ using an interaction kernel which has typically produced sensible results for hadron properties in-vacuum \cite{Roberts:2000aa, Maris:2003vk}.

We are thus presented with a quandary: how might one understand and reconcile this marked contradiction between simple, but apparently robust chiral-model predictions on one hand, and lQCD and well-constrained Dyson-Schwinger equation (DSE) studies on the other?


Resolving this predicament is the subject of our discussion.  We introduce the PNJL model in Sec.\,II, placing particular emphasis on the issue of ultraviolet regularization, which always plays a crucial role in any application of a contact interaction \cite{Farias:2005cr, Farias:2006cs, GutierrezGuerrero:2010md, Roberts:2011wy, Chen:2012txa, Farias:2014eca, Farias:2016let}.  Section~III updates DSE predictions for the phase diagram of QCD with $\mu_5\geq 0$.  That establishes a context for the discussion in Sec.\,IV, which canvasses the impact of different regularization schemes for the PNJL model on the existence, location and evolution of CEP$_5$ and CEP$_\chi$ in that model, with very instructive consequences.  We summarize and conclude in Sec.\,V.

\smallskip

\noindent\textbf{II.$\;$PNJL Model and Effective Potential}.
\label{pnjl}
The PNJL model for two flavors of equal-mass quarks may be defined by the following Lagrangian density:
\begin{align}
\nonumber
{\cal L}  & = \bar q (\gamma\cdot D + m )q \\
& \quad - G \left[ (\bar q q)^2 + (\bar q i \gamma_5 \tau q)^2 \right] + {\cal U}(\Phi,\bar\Phi;T)\,,
\label{eq1}
\end{align}
where: $m$ is the common current-quark mass;
$D_\mu = \partial_\mu +  i A_\mu$,
with $A_\mu(x) = g_s A_\mu^a \lambda^a/2$ describing the matrix-valued gluon field configuration appropriate to the model; $G$ is the four-fermion interaction strength; and ${\cal U}$ is a Polyakov-loop effective potential.

In general, the Polyakov loop is defined as the following matrix in color space, $SU_c(N_c=3)$:
\begin{equation}
L(x) = {\cal P}\exp[ -i \mbox{$\int_0^\beta \! dx_4\,A_4(x_4,\vec{x})$}]\,,
\end{equation}
where ${\cal P}$ is a path-ordering operator and $\beta = 1/T$.  However, in connection with the PNJL model, it is customary to define $L(x)$ in Polyakov gauge, which sets $A_4$ static and diagonal in color space, and require $L^\dagger = L$.  With these conventions \cite{Fukushima:2003fw}, the model's mean-field effective-potential can be written solely in terms of
\begin{equation}
\label{LPhi}
\Phi = \tfrac{1}{N_c}Tr_c\, L = \bar \Phi\,,
\end{equation}
which evolves with the intensive thermodynamic variables characterizing the medium.  The domain of confinement in the pure-gauge theory is expressed via $\Phi=0$, whereas $\Phi=1$ defines the deconfined domain.

In terms of the classical background field in Eq.\,\eqref{LPhi}, an efficacious representation of the Polyakov-loop effective potential is provided by \cite{Roessner:2006xn}:
\begin{align}
\nonumber
& \beta^4 {\cal U}(\bar{\Phi},\Phi;T)  = \beta^4 {\cal U}(\Phi;T) \\
& = -\tfrac{1}{2}a(T) \Phi^2 +b(T)\,\ln[1-6\,\Phi^2  +8 \Phi^3 -3\Phi^4],
\label{eq5}
\end{align}
with $(\bar t = T_0/T)$
\begin{equation}
a(\bar t)=a_0+a_1\bar t + a_2\bar t^2\,,\quad
b(\bar t)=b_3\bar t^3\,,
\label{eq6}
\end{equation}
where the parameters, listed in Table~\ref{bz1}, were chosen \cite{Roessner:2006xn} in order to reproduce lattice results for pure-gauge QCD thermodynamics and the $T$-dependence of the Polyakov loop.  Following Ref.\,\cite{Ruggieri:2011xc}, however, the value of $T_0$ is adjusted to account for the presence of dynamical quarks.

\begin{table}[t]
\caption{\label{bz1}
Parameter values used herein to define the PNJL model.
\emph{Upper panel} -- Polyakov-loop potential, Eqs.\,\eqref{eq5}, \eqref{eq6} \cite{Roessner:2006xn, Sakai:2010rp}.
\emph{Lower panel} -- NJL part of the Lagrangian density, Eqs.\,\eqref{eq1}, \eqref{Gchoice} \cite{Sakai:2010rp}, with dimensioned quantities in MeV.
}
\begin{center}
\begin{tabular*}
{\hsize}
{
c@{\extracolsep{0ptplus1fil}}
c@{\extracolsep{0ptplus1fil}}
c@{\extracolsep{0ptplus1fil}}
c@{\extracolsep{0ptplus1fil}}
c@{\extracolsep{0ptplus1fil}}}\hline
$a_0$&$a_1$&$a_2$&$b_3$&$T_0$\\
\hline
3.51&-2.47&15.2&-1.75&190 \\\hline
\end{tabular*}
\medskip

\begin{tabular*}
{\hsize}
{
c@{\extracolsep{0ptplus1fil}}
c@{\extracolsep{0ptplus1fil}}
c@{\extracolsep{0ptplus1fil}}
c@{\extracolsep{0ptplus1fil}}
c@{\extracolsep{0ptplus1fil}}}\hline
$m$ &$\Lambda$  &$g \Lambda^2$ &$\alpha_1$ &$ \alpha_2$\\
\hline
5.5 & 631.5 & $2.2 $ & $0.2$ &0.2\\\hline
\end{tabular*}
\end{center}
\end{table}

It is appropriate at this point to reflect upon the four-fermion coupling, $G$, in Eq.\,\eqref{eq1}, which is supposed to contain information about gauge-sector dynamics.  Since that dynamics is also expressed in $\Phi$, it might be imagined that a realistic model would replace $G\to G(\Phi)$.  Naturally, however, any such statement introduces additional model dependence.  Herein, we therefore explore two possibilities, \emph{viz}.\
\begin{subequations}
\label{Gchoice}
\begin{align}
\label{Ggconstant}
& \mbox{Ref.\,\cite{Fukushima:2003fw}}: \;
& G=g = \mbox{constant, \rule{3.3em}{0ex}}\\
& \mbox{Refs.\,\cite{Ruggieri:2011xc, Gatto:2011wc}}: \;
& G = g [ 1-\alpha_1 \Phi^2 - 2 \alpha_2 \Phi^3 ]\,,
\label{eq:Run}
\end{align}
\end{subequations}
with the parameters in Table~\ref{bz1} chosen such that additional aspects of the PNJL model are consistent with simulations of lQCD \cite{Sakai:2010rp}.

Finally, in order to study the interplay between $T$, and regular and chiral chemical potentials, we define the action with an upper bound $\beta$ on the $dx_4$ integral and add the following term to Eq.\,\eqref{eq1}:
\begin{equation}
\label{lagrmu5}
-\bar q \gamma_4[ \mu + \mu_5 \gamma_5 ] q \,.
\end{equation}

Adopting the mean-field approximation, one obtains the following effective potential for the PNJL model we have described \cite{Ruggieri:2011xc}:
\begin{align}
\nonumber
& \Omega = \Omega(M,\Phi;T,\mu,\mu_5)\\
\nonumber
& = {\cal U}(\Phi;T) + \frac{(M-m)^2}{4 G} - 2 N_c \sum_{s=\pm 1}\int\frac{d^3\vec{p}}{(2\pi)^3} \, \omega_s \\
& \quad - \frac{2}{\beta}  \sum_{s=\pm 1}\int\frac{d^3\vec{p}}{(2\pi)^3} \ln [ {\cal F}_+ \, {\cal F}_-]\,,
\label{Omega}
\end{align}
where $M$ is the DCSB-induced mass gap,
\begin{align}
\omega_s & = \sqrt{(s |\vec{p}|-\mu_5)^2 + M^2}\,,\\
{\cal F}_\pm & = 1 + 3 \Phi [{\rm e}^{-\beta \omega_s^\pm }
+  {\rm e}^{-2\beta\omega_s^\pm}] +  {\rm e}^{-3\beta\omega_s^\pm }\,,
\label{Fpm}
\end{align}
$\omega_s^\pm = \omega_s \pm \mu$.  At this point one can determine the evolution of the quark mass-gap with intensive parameters via simultaneous solution of the extremal conditions:
\begin{equation}
\frac{\partial \Omega}{\partial M} = 0 = \frac{\partial \Omega}{\partial \Phi}\,.
\end{equation}

It is worth noting that $s$ in Eqs.\,\eqref{Omega} -- \eqref{Fpm} is a chirality label, the sum over which appears owing to the presence of $\mu_5$ in the model.  Furthermore, the coupling between quarks and the Polyakov loop is prominently expressed through ${\cal F}_\pm$ in Eq.\,\eqref{Fpm}: in the gauge-confined phase, $\Phi=0$ and one has a standard NJL-model effective-potential; but for $\Phi\neq 0$, $\Omega$ contains couplings $\sim \Phi {\rm e}^{-\beta M}$, and consequently the deconfinement transition encoded in the Polyakov-loop can influence the chiral transition, expressed in the behavior of the quark mass-gap.

\begin{figure}[t]
\includegraphics[width=0.38\textwidth]{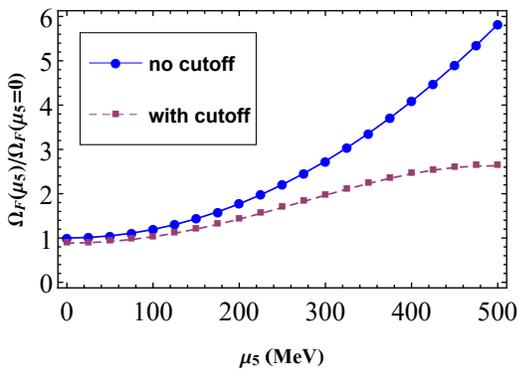}
\caption{Comparison between $\Omega_{\cal F}^{\Lambda}$ (dashed, purple curve) and $\Omega_{\cal F}^{\infty}$ (solid blue curve), evaluated with $(T=0.1,\mu=0.2)\,$GeV.  This comparison is not qualitatively sensitive to the precise values of $(T,\mu)$; and similar differences are also evident if one chooses $T$ or $\mu$ as the independent variable.  \label{diff}}
\end{figure}

Hitherto we have not explicitly addressed the question of regularization for the PNJL model.  The last term in the second line of Eq.\,\eqref{Omega},
\begin{equation}
\label{OmegaV}
\Omega_V = 2 N_c \sum_{s=\pm 1}\int\frac{d^3\vec{p}}{(2\pi)^3} \, \omega_s \,,
\end{equation}
is plainly divergent so that $\Omega$ is meaningless as written.  A regularization procedure must be introduced.  We employ a hard cutoff, \emph{viz}.\ $\Lambda$ in the lower panel of Table~\ref{bz1}.  Using that value, and $m$ and $g$ listed therewith, a good description of in-vacuum pion properties is obtained.

\begin{figure}[t]
\includegraphics[width=0.40\textwidth]{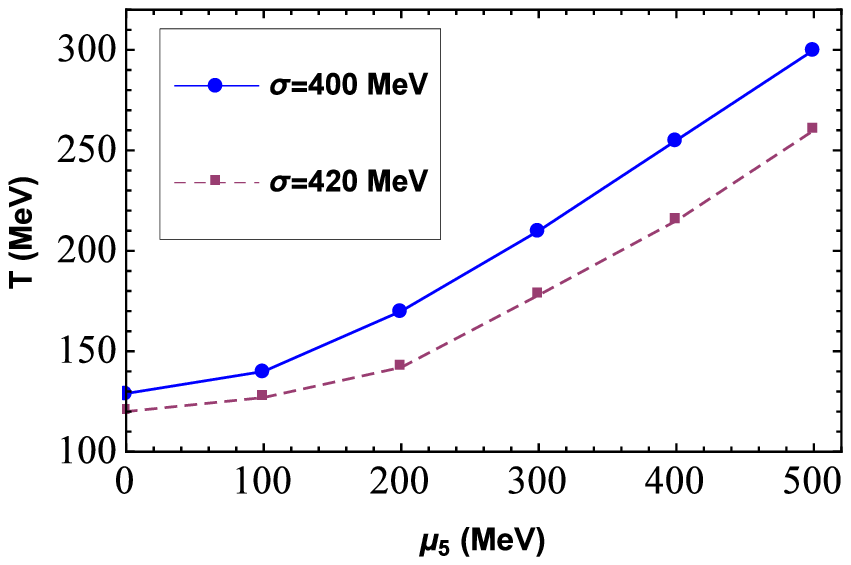}
\includegraphics[width=0.42\textwidth]{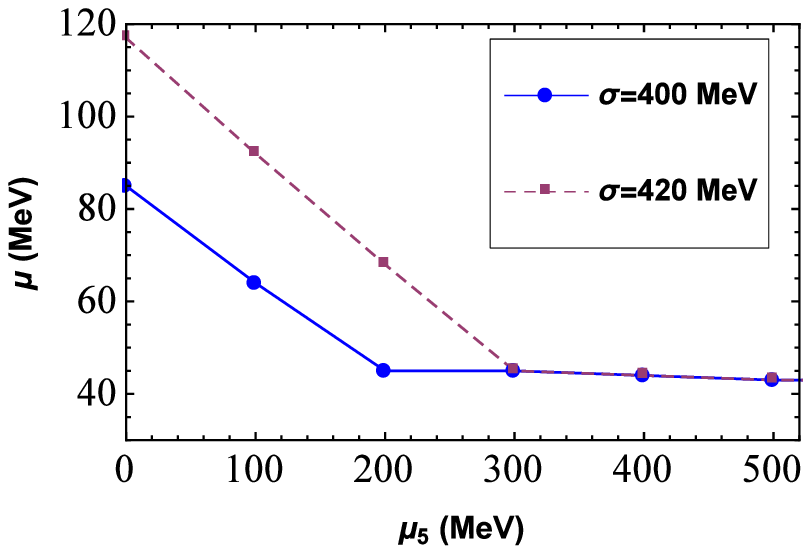}
\caption{\label{dsemu}
DSE predictions for the location of the critical endpoint associated with the chiral symmetry restoring transition, $(\mu(\mu_5),T(\mu_5))$: $T(\mu_5)$, upper panel; and $\mu(\mu_5)$ lower panel, computed with two different values of the mass-scale, $\sigma$, which determines the interaction strength in Eq.\,\eqref{mtgluon}.
All curves in both panels were computed as described in Ref.\,\cite{Xu:2015vna}.}
\end{figure}

The question which now arises, however, is what to do with the remaining integral in Eq.\,\eqref{Omega}?  The quantity
\begin{equation}
\label{OmegaF}
\Omega_{\cal F} = \frac{2}{\beta}  \sum_{s=\pm 1}\int\frac{d^3\vec{p}}{(2\pi)^3} \ln [ {\cal F}_+ \, {\cal F}_-]
\end{equation}
is finite, so a cutoff is not strictly necessary and none is used in Refs.\,\cite{Fukushima:2003fw, Roessner:2006xn, Ruggieri:2011xc, Fukushima:2010fe, Chernodub:2011fr, Gatto:2011wc}.  However, we question the spirit of this choice.

One justifies a regularization of $\Omega_V$, Eq.\,\eqref{OmegaV}, by observing that QCD is asymptotically free, so high-momentum modes should not materially influence nonperturbative strong interaction phenomena.  Indeed, the contact interaction itself can broadly be reconciled with QCD by imagining that the necessary regularization function is a coarse but useful representation of the transition between nonperturbative infrared dynamics, such as gluon mass-generation \cite{Cornwall:1981zr, Aguilar:2008xm, Boucaud:2011ug, Ayala:2012pb, Binosi:2014aea, Aguilar:2015bud}, and the domain of asymptotic freedom.  Adopting this perspective, it seems that internal consistency requires one to use a definition of $\Omega_{\cal F}$ which employs the same (or similar) cutoff used in connection with $\Omega_V$.

We will subsequently, therefore, compare results obtained with two procedures:
(\emph{i}) $(\Omega_V^\Lambda, \Omega_{\cal F}^{\Lambda})$, also explored in Refs.\,\cite{Hansen:2006ee, Costa:2009ae, Bratovic:2012qs},\footnote{A more sophisticated expression of this idea was exploited in Ref.\,\cite{Farias:2014eca} in order to reconcile NJL and lQCD results relating to the pseudocritical temperature in magnetized quark matter.}
and (\emph{ii}) $(\Omega_V^\Lambda, \Omega_{\cal F}^{\Lambda\to\infty})$.  The difference between these two definitions is depicted in Fig.\,\ref{diff}.  Given that the discrepancy grows with increasing $\mu_5$ (and $T$, $\mu$), it should not be surprising if considerable disparity were to emerge between the predictions made by (\emph{i}) and (\emph{ii}) concerning the existence and location of CEP$_{\chi,5}$, which, if at all, are likely to be found at larger values of the intensive parameters.

\smallskip

\noindent\textbf{III.$\;$DSE Predictions}.
\label{DSEresults}
As a prelude to detailing results obtained with the PNJL model, we recapitulate and update predictions for the location of the critical endpoint, $(\mu(\mu_5),T(\mu_5))$, associated with the chiral symmetry restoring transition, which have been obtained using DSE methods.  In this, we follow Ref.\,\cite{Xu:2015vna}, using the rainbow-ladder truncation \cite{Binosi:2016rxz} of the dressed-quark gap equation with the interaction in Ref.\,\cite{Qin:2010nq}:
\begin{equation}
\label{mtgluon}
g^2D_{\mu\nu}(k_{n}) = D_0 \frac{4\pi^2} {\sigma^6} k_{n}^2e^{-k_{n}^2/\sigma^2},
\end{equation}
where $D_0=(0.96\,$GeV$)^2$, $k_{n}=(\vec{k},\omega_{n})$, and $\omega_n=2n\pi T$ is a boson Matsubara frequency.  Here and in the following we locate the CEP by studying the behaviour of the chiral susceptibility, $\chi_M$, defined via the dressed-quark mass-function \cite{Blaschke:1998mp, Holl:1998qs}: the CEP is positioned at that set of intensive parameters for which $[1/\chi_M]\to 0$  \cite{Qin:2010nq}.

\begin{figure}[t]
\centerline{\includegraphics[width=0.38\textwidth]{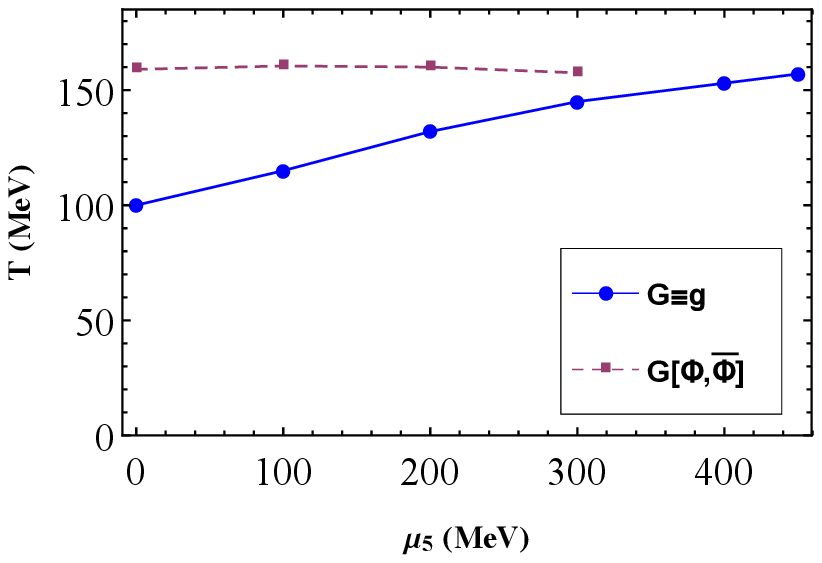}}
\centerline{\includegraphics[width=0.38\textwidth]{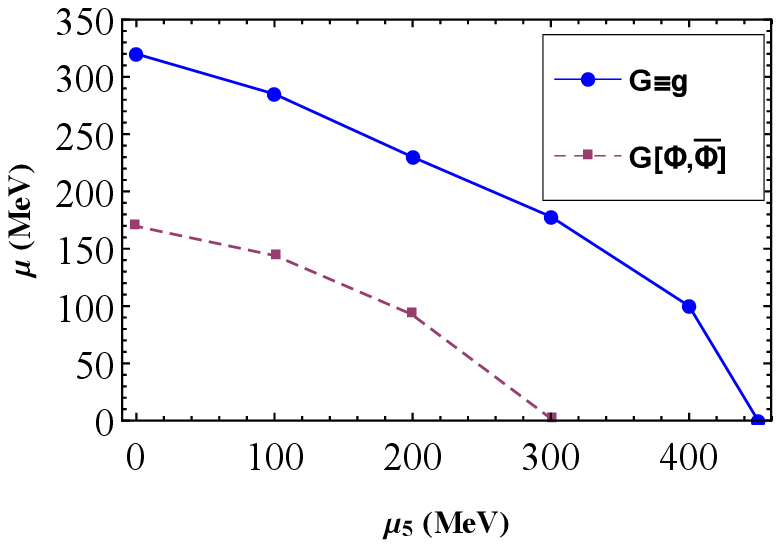}}
\caption{\label{cep1}
Location of the critical endpoint associated with the chiral symmetry restoring transition, $(\mu(\mu_5),T(\mu_5))$, computed using the PNJL model defined with $\Omega_{\cal F}^{\infty}$: $T(\mu_5)$, upper panel; and $\mu(\mu_5)$ lower panel.  The dashed curves are obtained using a constant NJL coupling, Eq.\,\eqref{Ggconstant} and the solid curves with a $\Phi$-dependent coupling, Eq.\,\eqref{eq:Run}.
}
\end{figure}

The results of this analysis, obtained with current-quark mass $m=5\,$MeV, are depicted in Fig.\,\ref{dsemu}.   We used two values of the strength-parameter, $\sigma$: in the limit $\sigma \to 0$, the interaction approaches a $\delta$-function \cite{Munczek:1983dx}.  Plainly, the temperature associated with the critical endpoint increases with $\mu_5$; but, although the correlated chemical potential does initially decrease with $\mu_5$, it fails to reach $\mu=0$ and hence there is no CEP$_5$.
The DSE predictions are evidently in qualitative agreement with those obtained using lQCD \cite{Yamamoto:2011gk, Braguta:2015zta, Braguta:2015owi}; but therefore differ markedly from the PNJL model results \cite{Ruggieri:2011xc, Fukushima:2010fe, Chernodub:2011fr, Gatto:2011wc}.

\begin{figure}[t]
\centerline{\includegraphics[width=0.39\textwidth]{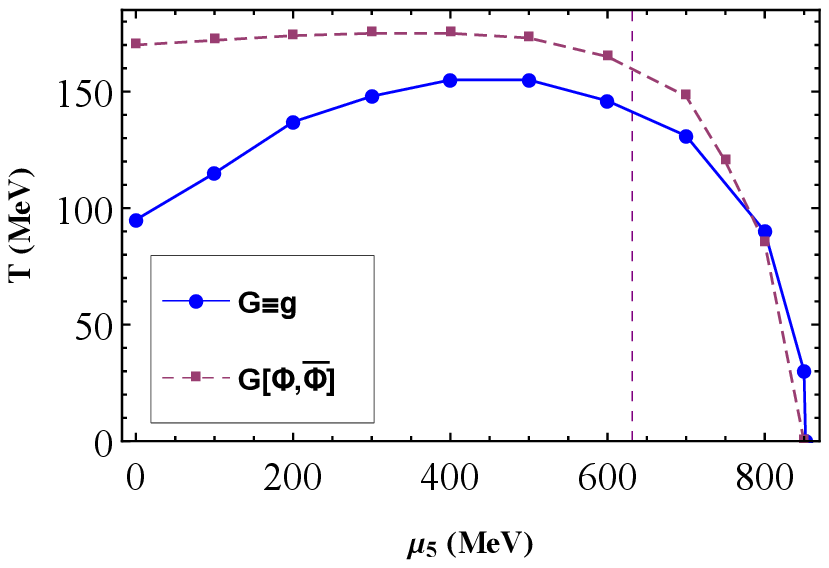}}
\centerline{\includegraphics[width=0.39\textwidth]{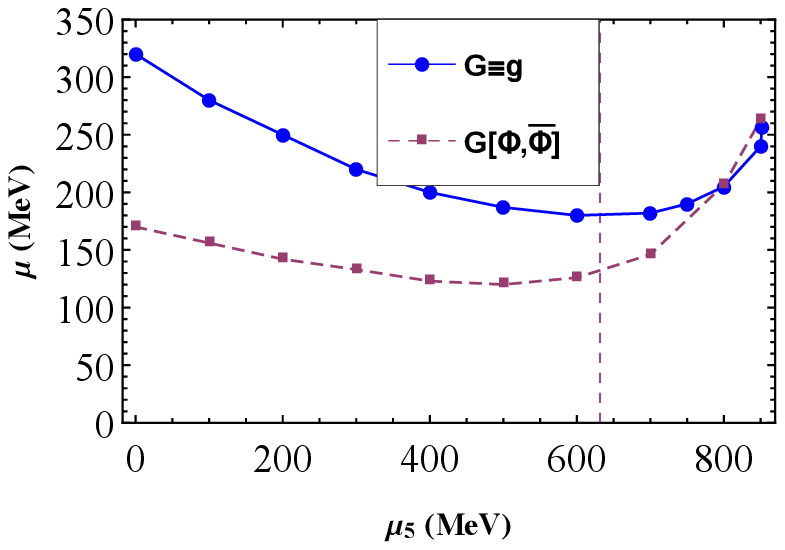}}
\caption{\label{cep3}
Location of the critical endpoint associated with the chiral symmetry restoring transition, $(T(\mu_5),\mu(\mu_5))$, computed using the PNJL model defined with $\Omega_{\cal F}^{\Lambda}$: $T(\mu_5)$, upper panel; and $\mu(\mu_5)$ lower panel.  The dashed curves are obtained using a constant NJL coupling, Eq.\,\eqref{Ggconstant} and the solid curves with a $\Phi$-dependent coupling, Eq.\,\eqref{eq:Run}.  The vertical lines mark the point $\mu_5=\Lambda$, \emph{viz}.\ the upper boundary for any sensible interpretation of the model's results.}
\end{figure}

The lower panel of Fig.\,\ref{dsemu} exhibits some curious features.  First, with decreasing $\sigma$, the value of $\mu$ at the CEP in the $(\mu,\mu_5,T)$-hyperplane decreases on a measurable domain containing $\mu_5=0$.  This is explained by the fact that in the limit $\sigma\to 0$, CEP$_\chi$ lies at $\mu=0$ for $m=0$ \cite{Blaschke:1997bj}.
The second curious feature is that for each value of $\sigma$ there is a critical value of $\mu_5=\mu_5^i(\sigma)$ such that $\forall \mu_5>\mu_5^i(\sigma)$ the value of $\mu$ associated with the critical endpoint in the $(\mu,\mu_5,T)$-hyperplane is independent of $\mu_5$.  We have established that this constant value of $\mu=\mu^i$ is determined by the current-quark mass, $\mu^i=\mu^i(m)$: with $m=5\,$MeV, $\mu^i \approx 40\,$MeV; and $\mu^i\approx 90\,$MeV for $m=15\,$MeV.  Accordingly, $\forall \mu < \mu_i(m\neq 0)$, the chiral transition is a crossover.  The existence and evolution of $\mu^i(m)$ can be understood by exposing the impact of $m$ on the analytic structure of the dressed-quark propagator \cite{Zong:2005mm, Chen:2008zr}; and in this, too, the algebraic model of Ref.\,\cite{Blaschke:1997bj} can be used profitably.

\smallskip

\noindent\textbf{IV.$\;$PNJL Model: Results and Remarks}.
\label{puzzle}
We turn now to a discussion of results obtained using the PNJL model.
In Fig.\,\ref{cep1} we depict trajectories of the critical endpoint for the chiral symmetry restoring transition obtained when the PNJL model is defined using $\Omega_{\cal F}^{\infty}$, \emph{i.e}.\ eschewing a limitation on the high-momentum modes in the last term of the effective potential \cite{Fukushima:2003fw, Roessner:2006xn, Ruggieri:2011xc, Fukushima:2010fe, Chernodub:2011fr, Gatto:2011wc}, and with both choices of the NJL four-fermion coupling identified in Eqs.\,\eqref{Gchoice}.  Evidently, irrespective of the latter choice, and  in contradistinction to lQCD and DSE results, a CEP$_5$ exists.  On the other hand, it is apparent that if one uses $G=g=\,$constant, Eq.\,\eqref{Ggconstant}, then the temperature associated with the critical endpoint for the chiral symmetry restoring transition does increase with $\mu_5$, in agreement with lQCD and DSE analyses.  This, however, is not the definition employed in Refs.\,\cite{Sakai:2010rp, Ruggieri:2011xc, Gatto:2011wc}: they employed Eq.\,\eqref{eq:Run}.

\begin{figure}[t]
\centerline{\includegraphics[width=0.42\textwidth]{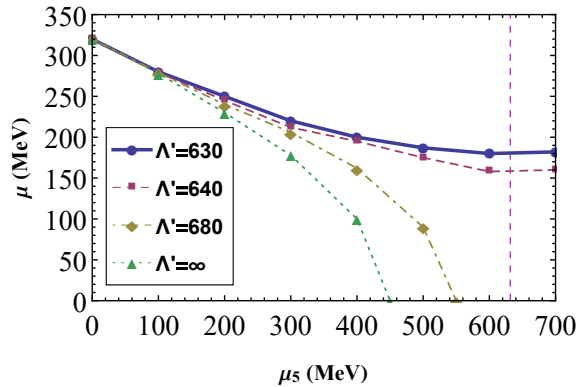}}
\caption{\label{mumu5Lambda}
Trajectories of the chemical potential associated with the chiral transition's critical endpoint, $\mu(\mu_5)$, computed with $G=g=\,$constant, Eq.\,\eqref{Ggconstant}, in PNJL models defined using various different values for the cutoff on the last term in the effective potential, Eq.\,\eqref{OmegaF}: $\Omega_{\cal F}^{\Lambda^\prime}$: $\Lambda/\Lambda^\prime = 0,0.92,0.99,1$.
}
\end{figure}

In Fig.\,\ref{cep3} we depict trajectories of the critical endpoint obtained with the $\Omega_{\cal F}^{\Lambda}\,$-PNJL model, \emph{i.e}.\ produced by introducing a physically-motivated cutoff on the high-momentum modes in the last term of the effective potential, and with both choices of the four-fermion coupling identified in Eqs.\,\eqref{Gchoice}.
We observe first that when using $\Omega_{\cal F}^{\Lambda}$, one should restrict the domain of model applicability to values of the intensive parameters which lie below the cutoff, \emph{i.e}.\ $\mu_5\lesssim \Lambda$ in the present instance: results on this domain can reasonably be expected to be sensible.   (This limitation can be eliminated by using a better regularization scheme \cite{Farias:2005cr, Farias:2006cs, GutierrezGuerrero:2010md, Roberts:2011wy, Chen:2012txa, Farias:2014eca, Farias:2016let}; but such improvements have no material implications for the present discussion.)
Bearing the restriction in mind, it then becomes apparent that the $\Omega_{\cal F}^{\Lambda}\,$-defined PNJL-model predictions obtained with $G=g=\,$constant, Eq.\,\eqref{Ggconstant}, are qualitatively in agreement with lQCD and DSE results: the temperature associated with the critical endpoint of the chiral transition increases with $\mu_5$ and there is no CEP$_5$.

Evidently, as anticipated in the conclusion to Sec.\,II, 
the differences highlighted by Fig.\,\ref{diff} have a significant impact on the PNJL model's qualitative features.  This is illustrated further by Fig.\,\ref{mumu5Lambda}, which shows that there is a critical value for the physically motivated cutoff employed in connection with $\Omega_{\cal F}$, $\Lambda^c$, such that no CEP$_5$ exists for any $\Lambda < \Lambda^c$.  The result $\Lambda^c \approx \Lambda$ highlights again the importance of an internally-consistent limitation on the contribution to the effective potential from high-momentum quark modes.

\smallskip

\noindent\textbf{V.$\;$Conclusion}.
\label{epilogue}
%
We set out to reconcile marked differences between predictions made by a class Polyakov-loop-modified Nambu--Jona-Lasinio (PNJL) models for the behavior of the chiral symmetry restoring transition in the presence of a chiral chemical potential, $\mu_5$, and those produced by lattice-QCD (lQCD) and Dyson-Schwinger equation (DSE) studies which provide a good description of low-energy $\pi$- and $\rho$-meson properties.
We found that the resolution lies with the nature of the regularization scheme employed to define the PNJL models.
All approaches are in qualitative agreement [Fig.\,\ref{cep3} \emph{cf}.\ Fig.\,\ref{dsemu}] so long as both (\emph{i}) a regularization procedure is employed to suppress high-momentum quark-modes in all terms that appear in the definition of the effective potential connected with the PNJL models, which seems a physically sensible requirement, and (\emph{ii}) the four-fermion coupling in those models does not react very strongly to changes in the mean-field that is assumed to mock-up Polyakov loop dynamics.
If one accepts this as providing the more realistic definition of PNJL models, then, on their domain of validity, the model predictions agree with those made by lQCD and DSE studies; and consequently there is no longer reason to expect that simulations of lQCD with $\mu_5>0$ will shed any light on the existence and location of a critical endpoint in the phase diagram of QCD in the $(T,\mu)$-plane.


\smallskip

\noindent\textbf{Acknowledgments}.
This work is supported by:
National Natural Science Foundation of China (under grant nos.~11275097, 11475085, and 11535005);
Jiangsu Planned Projects for Postdoctoral Research Funds (under grant no.~1402006C);
China Postdoctoral Science Foundation (under grant no.~2015M581765);
U.S.\ Department of Energy, Office of Science, Office of Nuclear Physics, under contract no.~DE-AC02-06CH11357;
and the Chinese Ministry of Education, under the \emph{International Distinguished Professor} programme.


\end{document}